\documentclass[aps,prd,twocolumn,superscriptaddress,amsfont,amssymb,amsmath,nofootinbib,showpacs,balancelastpage]{revtex4-1}
\usepackage{graphicx,longtable,natbib,mathrsfs,color}

\newcommand{\diff}{{\mathrm d}}

\newcommand{\LCDM}{$\Lambda$CDM }
\newcommand{\HI}{{H{\small\,I}}}

\begin{document}

%\addtolength{\hoffset}{-0.525cm}
%\addtolength{\textwidth}{1.05cm}
\title{Method for Direct Measurement of Cosmic Acceleration by 21-cm Absorption Systems}

%\author{Hao-Ran~Yu}\email{\url{haoran@cita.utoronto.ca}}\affiliation{Department of Astronomy, Beijing Normal University, Beijing 100875, China}
\author{Hao-Ran~Yu}\affiliation{Department of Astronomy, Beijing Normal University, Beijing 100875, China}\affiliation{Kavli Institute for Astronomy and Astrophysics, Peking University, Beijing 100871, China}
\author{Tong-Jie~Zhang}\email{\url{tjzhang@bnu.edu.cn}} \affiliation{Department of Astronomy, Beijing Normal University, Beijing 100875, China}
\author{Ue-Li~Pen}\email{\url{pen@cita.utoronto.ca}} \affiliation{Canadian Institute for Theoretical Astrophysics, University of Toronto, Toronto, M5S 3H8, Ontario, Canada}

\date{Received \today; published -- 00, 0000}

\begin{abstract}

So far there is only indirect evidence that the Universe is undergoing an accelerated expansion. The evidence for cosmic acceleration is based on the observation of different objects at different distances, and requires invoking the Copernican cosmological principle, and Einstein's equations of motion.  We examine the direct observability using recession velocity drifts (Sandage-Loeb effect) of 21cm hydrogen absorption systems in upcoming radio surveys.  This measures the change in velocity of the {\it same} objects separate by a time interval and is a model-independent measure of acceleration.  We forecast that for a CHIME-like survey with a decade time span, we can detect the acceleration of a $\Lambda$CDM Universe with $5\sigma$ confidence. This acceleration test requires modest data analysis and storage changes from the normal processing, and cannot be recovered retroactively.

\end{abstract}


\maketitle

\textit{Introduction.---}
One of the biggest mysteries in physics is the purported accelerated
expansion of the Universe.  This suggests that gravity on the largest
scales is not an attractive force, but for the Universe at large is
repulsive.  Strong mysteries require strong and preferably direct
evidence, and to date the inference of acceleration is indirect.
Direct evidence would require the measurement of velocities of objects
at cosmological distances at different epochs, and determining an
increase in recession velocity for the {\it same} objects.

Current indirect inferences require the application of
Einstein's equation combined with a Copernican principle.  The
Copernican principle underlies the assumption of large scale
homogeneity of the cosmos, which enables the general solution of the
complex, non-linear Einstein equations.  The indirect inferences
include dynamical measurements of the gravitational evolution of
structure\citep{1995Natur.377..600O,2013arXiv1303.5079P_comment}, or
kinematic measurements using luminosity or angular diameter distances
\citep{1998AJ....116.1009R,1999ApJ...517..565P,2000Natur.404..955D,2013arXiv1303.5076P,2005ApJ...633..560E}.

They are indirect because they measure different objects at multiple
distances (redshifts) in the Universe at an instant of time on earth,
and the acceleration is inferred by assuming a homogeneous
cosmological model and an equation of motion. The direct
model-independent physical acceleration is by definition the velocity
change over a time interval between two measurements. A direct probe
of the cosmic acceleration is the Sandage-Loeb (SL)
effect \cite{1962ApJ...136..319S,1998ApJ...499L.111L}, suggested by
Sandage to measure change of redshift of galaxies, and by Loeb to
measure the drift in the Ly$\alpha$ forest. The latter proposal
motivates the construction of the high-resolution CODEX\footnote{\url{http://www.iac.es/proyecto/codex/}} (COsmic Dynamics and EXo-earth
experiment) 
 spectrogragh on E-ELT\footnote{\url{http://www.eso.org/public/teles-instr/e-elt.html}} (European
Extremely Large Telescope), to measure the
precise redshift of Ly$\alpha$ over a two
decade interval.
The CODEX group
provide a full design for observing the SL signal and a prediction
of the statistical error of the SL
signal \citep{2008MNRAS.386.1192L}. It is also shown that the effect
has the potential to better constrain dark energy
\citep{2007PhRvD..75f2001C,2013PhRvD..88b3003L,2013arXiv1311.1583Y}.  This ambitious study
primarily searches for cosmic deceleration and jerk at redshift $z \gtrsim
2$, and is not sensitive to direct acceleration measurements.  For
a direct measurement of acceleration, a lower redshift is preferable,
when the Universe is actually accelerating.

In the radio domain, the 21cm line in the hydrogen absorption systems
are the most promising candidates to detect the acceleration. In
damped Ly$\alpha$ absorption systems (DLA), the radio spectrum is also
substantially absorbed by the neutral hydrogen (\HI) hyperfine
structure which has the rest frame wavelength of approximately 21cm or
frequency of 1420 MHz.  DLA's are defined as having \HI\ column
density greater than $2\times 10^{20}$ cm$^{-2}$.  This density
results in substantial self-shielding, and empirically is also the
minimum column to house cold 21cm absorbing gas in the cold neutral
medium (CNM).  Treated as 21cm hydrogen absorption systems, the
measurements are readily accessible over the cosmic accelerating
redshift using ground based radio telescopes.  Their advantages
include: (1) the intrinsic line width is narrow because absorption is
dominated by cold absorbers at $T<80$ K. (2) collecting area is cheap
in the redshifted 21cm band. (3) a new generation of telescopes is
being constructed in the relevant bands.  The best direct constraint
on the acceleration of the Universe to date is based on observing
these systems \cite{2012ApJ...761L..26D} over more than a
decade. These errors are still three orders of magnitude larger than a
prediction from a \LCDM (cold dark matter with a cosmological constant $\Lambda$) Universe and overwhelm the real signal.

Large numbers of 21cm absorption systems in the purported accelerating
regime of the Universe will be detected by wide-sky
radio surveys like
PARKES\footnote{phased array feed project on parkes, \url{http://www.atnf.csiro.au/research/multibeam/}},
under-construction
ASKAP\cite{2011ApJ...742...60D},
CHIME\footnote{\url{http://CHIMExperiment.ca}},
GBT-multibeam\footnote{\url{http://gbt800mb.pbworks.com/}}
and proposed
BAOBAB\citep{2013AJ....145...65P},
BAOradio\citep{2012CRPhy..13...46A},
BINGO\citep{2012arXiv1209.1041B},
CARPE\footnote{\url{http://www.phys.washington.edu/users/mmorales/carpe/}},
MeerKAT\footnote{\url{http://www.ska.ac.za/meerkat/}},
SKA\footnote{\url{http://www.skatelescope.org}} and
Tianlai\footnote{\url{http://tianlai.bao.ac.cn}}
\citep{2012IJMPS..12..256C}, 
etc.
In the remainder of this letter we will estimate how this tiny SL effect
could be extracted statistically within a decade from
instruments that are being constructed or proposed, in many cases for
the purpose of more precise indirect measurements of dark energy.

\textit{Quantifying the direct acceleration.---}
The acceleration $\dot{v}$ of a given object at redshift $z$ is given by its redshift drift
\begin{equation}\label{eq.vdot}
	\dot{v}=c\dot{z}/(1+z),
\end{equation}
where over-dots denote derivatives respect to observation time $\diff/\diff t$ and $c$ is the speed of light. Here the redshift drift $\dot{z}$ is linearly related to the Hubble parameter $H(z)$ at the object's redshift $z$
\begin{equation}
	\dot{z}=(1+z)H_0-H(z),
\end{equation}
where Hubble constant $H_0$ is the current expansion rate $H(z=0)$ and
$H(z)$ is given by the specified cosmological model. In a concordance
\LCDM Universe for instance, objects with redshift $z\lesssim 2.5$ are
accelerating and their acceleration $\dot{v}$ is by order of 0.1 cm
s$^{-1}$ yr$^{-1}$. To measure this minuscule velocity difference one
needs to subtract off proper accelerations of the observer. It can be
precisely measured by pulsar timing in the Galactic frame \citep{2005AJ....130.1939Z,2012A&A...544A.135X} 
and by proper motion of the extra-galactic radio sources
in the cosmological frame \citep{2011A&A...529A..91T}. Even without the observations of the proper accelerations and without full-sky coverage, one can still solve the cosmic and local accelerations by a moment analysis over the sky.

%observer acceleration only enters as a higher order relativistic term.

\textit{Feasibility.---}
Assuming a \LCDM Universe, the 21cm absorption systems in the redshift range $0<z<2.5$ are accelerating, which corresponds to the frequency range from 1420 MHz to 406 MHz. CHIME ($0.8<z<2.5$) and PARKES ($0<z<1.0$) -like surverys are suitable to be devised to scan the radio sources in NRAO VLA Sky Survey (NVSS\footnote{\url{http://www.cv.nrao.edu/nvss/}}), looking for possible 21cm absorption systems. Here we discuss a decade-long-survey by a CHIME-like telescope as an example.

% number counts

\begin{figure}
	\centering
	\includegraphics[width=0.5\textwidth]{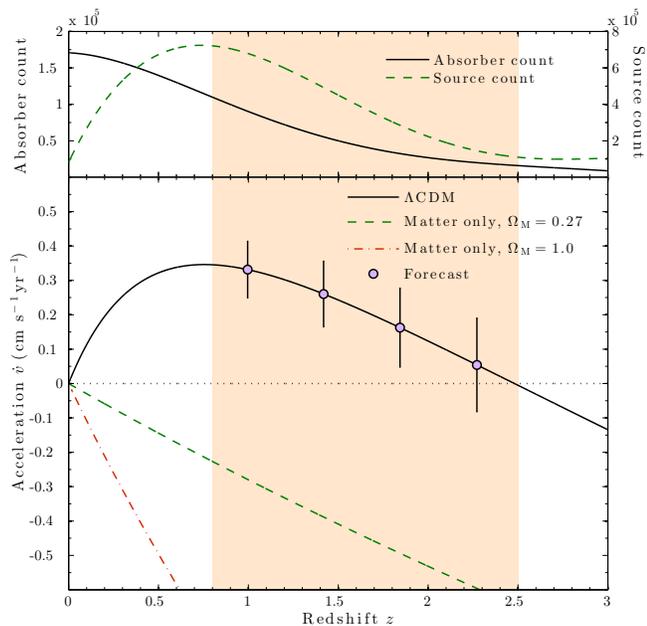}
\caption{(color online.) (\textit{top}:) Northern-hemisphere number counts per unit redshift ($\diff N/\diff z$) for 21cm absorption systems $f_{\rm 21cm}$ in black solid line, and NVSS radio sources $f_{\rm QSO}$ in (green) dashed line. (\textit{bottom}:) Velocity drift forecast by 21cm absorption systems by CHIME-like telescope assuming a concordance \LCDM Universe. The observation span is 10 years. The black solid line is the theoretical prediction by a \LCDM Universe, while the predictions of matter-only universes $\Omega_{\rm M}=0.27$, $\Omega_{\rm M}=1$, which are decelerating, are shown in (green) dashed line and (red) dash-dotted line.  CHIME's observable redshift range $0.8<z<2.5$ is shown by (orange-) shaded regions in both panels of the figure.}
	\label{fig:1}
\end{figure}

If the survey scans the northern-hemisphere of the sky, it is about $60\%$
area of NVSS coverage. NVSS contains a catalog of almost 2 million
discrete sources, with flux densities $F$ brighter than about 2.5 mJy at
1.4 GHz. The CHIME-like survey will be sensitive to absorbers in the 1.2 million NVSS
sources in its field of view. CHIME's frequency coverage requires
$\nu<800$ MHz, where the sources are typically brighter by
$\nu^{-0.7}$.  We choose a lower flux limit of 4.5 mJy, where the
source counts are well understood, and which is more than 17$\sigma$ above the thermal noise limit. For a rough estimation of the redshift distribution of the observed 21cm absorption systems, we may adopt a NVSS redshift
distribution $n_{_{\rm R}}(z)$ ({\it e.g.} eq(26) in
\cite{2010A&ARv..18....1D}) and neutral hydrogen cloud redshift
distribution $n_{_{\rm \HI}}(z)$ (the density of optical damped Ly$\alpha$
systems is 14\% per unit redshift \citep{2005ARA&A..43..861W}). If
those two components are uncorrelated and randomly distributed
throughout the sky, the resulting redshift distribution of 21cm
absorption systems will be $n_{_{\rm 21cm}}(z)=n_{_{\rm \HI}}(z)\int_{z}^{\infty} n_{_{\rm R}}(z')\diff z'$. We plot $n_{_{\rm 21cm}}(z)$ and $n_{_{\rm R}}(z)$ in the upper panel of Fig.\ref{fig:1}.

% sensitivity
The signal-to-noise ratio ($S/N$) equals to the flux $F$ of the radio
sources divided by the error of the measurement $\Delta F$.  For a
dual polarized system,
\begin{equation}\label{eq.SN}
	S/N=F/\Delta F=F\sqrt{2\Delta\nu\Delta t}/{\rm SEFD},
\end{equation}
where $\Delta\nu$, $\Delta t$ are the line width and integration time
respectively, and SEFD is the system equivalent flux density. For
CHIME, we adopt SEFD=25 Jy. 

The line width $\Delta\nu=u_{\rm width}\nu/c$, where $u_{\rm width}$
is the equivalent width of the \HI\ absorber, and $\nu$ is the frequency
of the absorber. The properties of 21cm absorbers are primarily derived from followup
studies of optical absorbers. 
The discovery rate from the cross correlation is a lower bound on the
expect number of absorbers, since high column density systems in the
CNM may systematically obscure potential background optical sources.
There are only 3 blind radio detections, and a survey may discover
more systems which are optically obscured.  They would likely be cold
and at high column density.
For sensitivity purposes, we treat all sources as $u_{\rm width}$ of 2 km s$^{-1}$ \citep{1982ApJ...259..495W,2005ARA&A..43..861W}.
In practice, sources typically have an equivalent $u_{\rm width}$ of
2 km s$^{-1}$, and an actual line structure which is varied and diverse, and beyond of the scope of this paper. Wider structures
potentially reduce the survey sensitivity. On the other hand, radio
searches will preferentially find cold and narrow line systems,
potentially increasing the sensitivity beyond the values adopted
here. Thus the line width $\Delta\nu\simeq 9.3$ kHz. The integration time $\Delta t=t_{\rm survey}\tau$, where $t_{\rm survey}$ is the total time survey
duration and $\tau=\lambda/2\pi D$ is the fractional time that an object on the celestial equator transits the field of view of the telescope\footnote{Objects at higher latitude have greater $\tau$. If objects are uniformly distributed on the sky and $\lambda\ll D$, we have a boost factor $b=\pi/2$ on integration time or $(S/N)^2$. We take this factor into account in the calculation.}. For CHIME, the diameter $D$ of each cylinder is 20 m and the observational wavelength $\lambda$ is 21cm$(1+z)$. For a ten-year survey ($t_{\rm survey}=3.16\times 10^8$ sec), $\Delta t\simeq 5.28\times 10^5(1+z)$ sec at the equator. Substituting $\Delta\nu$ and $\Delta 
t$ into Eq.(\ref{eq.SN}), we get $\Delta F=0.252(1+z)^{-1/2}$ mJy.

In order to construct the velocity and acceleration estimator we write $v=v_0+\dot{v}t+\eta$ and $\eta$ is Gaussian noise: $\left\langle \eta(t)\eta(t')\right\rangle=\sigma_\eta^2\delta(t-t')$. We have the velocity estimator $E_v=(1/t_{\rm survey})\int v\diff t$ and acceleration estimator
$E_{\dot{v}}=(12/t_{\rm survey}^3)\int vt\diff t$ such that $\left\langle E_v\right\rangle=v$ and $\left\langle E_{\dot{v}}\right\rangle=\dot{v}$ (integrals go from $-t_{\rm survey}/2$ to $t_{\rm survey}/2$). Their errors can be estimated by $\sigma_v^2=\left\langle E_v^2\right\rangle-\left\langle E_v\right\rangle^2$ and $\sigma_{\dot{v}}^2=\left\langle E_{\dot{v}}^2\right\rangle-\left\langle E_{\dot{v}}\right\rangle^2$.
By integration we can get $\sigma_v=\sigma_\eta t_{\rm survey}^{-1/2}$ and $\sigma_{\dot{v}}=2\sqrt{3}\sigma_\eta t_{\rm survey}^{-3/2}$, meaning that $\sigma_{\dot{v}}=2\sqrt{3}\sigma_v t_{\rm survey}^{-1}$: error is enlarged as we are requiring the time derivative information. The forecast of $\sigma_v$ also depends on various of factors like the actual line structure and sampling quantization. For a rough estimation, treating line structure as Gaussian with standard deviation $\sigma$, a 16-level quantization (with frequency bin 0.34$\sigma$) gives at least 98.8\% quantization efficiency \citep{2001isra.book.....T}, and thus 340 m s$^{-1}$ sampling will be enough. Other effects will become limiting factors as the quantization efficiency approaches unity when we do a even finer frequency sampling. According to the central limit theorem, The error of the velocity estimator $\sigma_v$ is given by $u_{\rm width}(S/N)_{\rm bin}^{-1 }$ so we can as well forecast $\sigma_{\dot{v}}$, where in general,
\begin{equation}\label{eq.SN_eff}
	(S/N)^2_{\rm bin}=\iint_{F_{\rm min}} ^{\infty} \left( \frac{bF^2}{\Delta F^2}
	n_{_{\rm \HI}}(z) \int_z^{\infty} n_{_{\rm R}}(F,z') \diff z' \right) \diff F\diff z,
\end{equation}
is the cumulative $(S/N)^2$ in one redshift bin. In relevant frequency bands, all $\diff n_{_{\rm R}}/\diff F$ have similar profile \citep{1984ApJ...287..461C} and using any one of those gives nearly the same result (we use $\diff n_{_{\rm R}}/\diff F$ at 0.61 GHz \citep{1984ApJ...287..461C}, as it is the
main contribution of background sources), so we do not assume any redshift dependence on the flux distribution. Thus the inner integration in Eq.(\ref{eq.SN_eff}) can be written as $\int_{z}^{\infty} n_{_{\rm R}}(z')\diff z'$. 

Binning all the data into four redshift bins from redshift 0.8 to 2.5, we plot the binned velocity drift forecast
$\dot{v}(z)$ with error bars in the lower panel of Fig.\ref{fig:1}. 

For the assumed \LCDM Universe, we can detect the acceleration with
$5.1\sigma$ confidence.  If taken as an indirect test of acceleration,
this same data in combination with Einstein's equations and
homogeneity could be used to exclude matter-only universes with
$\Omega_{\rm M}=0.27$ and $\Omega_{\rm M}=1$ with $12.5\sigma$ and
$29.4\sigma$ confidence respectively.

\textit{Discussion.---}
CHIME and other experiments are constructed to make precision indirect
measurements of dark energy.  Modest real-time analysis changes could
allow the {\it direct} detection of cosmic acceleration.  The data
would need to be recorded at sufficient spectral resolution, better
than 340 m s$^{-1}$, corresponding to 800 Hz, which is not
originally planned.  The frequency channelization cost is FFT based,
and increasing from a spectral resolution of 300 to $10^5$ doubles the
FFT cost.  Spatial computational costs are in principle unchanged, but
there could be additional overhead costs for the larger resulting data
sets. Frequency stability of $10^{-11}$ over a decade is required, which is
straightforward on decade time scales with GPS-rubidium clocks, but
also needs to be built into the system from the beginning.
Foreground contamination 
would not be a problem in these very narrow frequency bands: one
expects the spatial-frequency mixing of foregrounds for the filled
aperture experiment to be $(\delta \lambda/\lambda)^2$, which leaves the
foregrounds far below the thermal noise.  

The substantial improvement from previous 
estimates \cite{2012ApJ...761L..26D} arises from the persistent daily
observations of every system, and the 10,000 fold increase in number
of targets from the rapid all sky survey.  We have applied a
conservative cut on source detections of more than 17$\sigma$.  The
number of targets could be larger if compensation for false detection
rates is allowed. Due to the paucity of known 21cm absorbers, and
absence of accurate blind surveys, the actual 21cm detection rate
could be different from our assumptions \citep{2010MNRAS.401..743A}. Astronomical complexities,
including multiple absorption features within systems, variations in
optical depths and line width, are not well characterized and have
been neglected.  This could lead to sensitivity changes, either
positive or negative.  We stress that the incremental effort needed
for this experiment is minimal, and well worth the effect, even if
only to characterize the large number of absorption systems.

Detection of the acceleration by promising 21cm absorption systems
needs redshift below 2.5 which corresponds to the frequency from 1420
MHz to 406 MHz. Velocity drift data on frequencies lower than 406 MHz
($z>2.5$) will show decelerations and are no longer the most direct
evidence of accelerating expansion. If CHIME is scheduled for completion
in 2015, a ten-year campaign could result in direct detections in
2025. The above estimation has substantial room for a further
improvement as we can also include the southern-hemisphere through
SKA and potential southern hemisphere CHIME-like telescope.  Completing
the entire sky completely removes the 
acceleration bias in Galactic and cosmological frame. Moreover, by
using all frequencies up to 1420 MHz, it will also broaden the
redshift range, as there are still lots of absorber counts at
$0<z<0.8$ with obvious acceleration (see Fig.\ref{fig:1}). 
If combined with a southern hemisphere ``CHIME''-like
experiment with a decade cadence, the cosmic acceleration measurement
would be improved to a $\sim 8\sigma$ confidence level.

Our proposed experiment observes only the velocity changes of {\it same} single objects over a time interval, and thus the acceleration measure is geometrical and does not require any assumptions of homogeneity, isotropy and Einstein equations. However, in order to test the dark energy or Lemaitre-Tolman-Bondi (LTB) models, we still need to include isotropy, homogeneity and metric assumptions \citep{2008PhRvL.100s1303U,2012PhR...521...95Q,2013PhR...530...87W}. On the other hand, without these assumptions, adequate accuracy of the measurement also enables us to test any anisotropic cosmic acceleration or inhomogeneity of the Universe non-parametrically.

%, and disfavor matter-only universes 
% with $\Omega_{\rm M}=0.27$ and $\Omega_{\rm M}=1$ with $23\sigma$ and
% $55\sigma$ confidence respectively.

%That might be by far the most predictable and realistic experiment to prove the cosmic acceleration in the {\it near} future. 

\textit{Conclusions.---}
We have estimated the sensitivity of upcoming radio experiments to a
direct cosmic acceleration search.  We conclude that this detection
may be possible with a CHIME-like and subsequent telescopes, if appropriate
real-time data processing modifications were made.  A direct detection
of cosmic acceleration bypasses Copernican cosmological
principle and Einstein Equation assumptions normally
required to infer the most mysterious property of the Universe:
acceleration. 

\textit{Acknowledgments.---}
We thank Avi Loeb, Keith Vanderlinde, Nissim Kanekar, Jeremy Darling and Mark Halpern for helpful comments and discussions. This work was supported by the National Science Foundation of China (Grants No. 11173006), the Ministry of Science and Technology National Basic Science program (project 973) under grant No. 2012CB821804, and the Fundamental Research Funds for the Central Universities.

\bibliographystyle{h-physrev3}
\bibliography{haoran_ref}

\end{document}